\documentclass[a4paper,scale=0.8,onecolumn,nobibnotes,nofootinbib]{revtex4}
\usepackage{enumitem}
\usepackage{amsmath,amssymb}
\usepackage{graphicx} 
\usepackage{color}
\usepackage{epstopdf}
\usepackage{hyperref}
\usepackage{geometry}
\usepackage{subfigure}
\usepackage{float}
\usepackage{comment}

\usepackage{ulem}
\usepackage{booktabs}

\hypersetup{
    colorlinks=true,
    citecolor=blue,
    linkcolor=red,
    filecolor=magenta,
    urlcolor=cyan,}

\newcommand{\be}{\begin{equation}}
\newcommand{\ee}{\end{equation}}

\begin{document}

\title{
GUP-corrected black holes: Thermodynamic properties, evaporation time and shadow constraint from EHT Observations of M87* and Sgr A*
}
\author{H. Chen\footnote{haochen1249@yeah.net (corresponding author)}$^{1}$,S.-H. Dong\footnote{dongsh2@yahoo.com}$^{2,3}$
 E. Maghsoodi\footnote{e.maghsoodi184@gmail.com}$^{4}$, S. Hassanabadi\footnote{s.hassanabadi@yahoo.com}$^{5}$,J. K\v{r}i\v{z}\footnote{jan.kriz@uhk.cz }$^{5}$,
 S. Zare\footnote{szare@uva.es}$^{6}$, and
 H. Hassanabadi\footnote{h.hasanabadi@shahroodut.ac.ir}$^{5}$ }
\affiliation{$^{1}$ School of Physics and Electronic Science, Zunyi Normal University, Zunyi 563006, China.\\
$^{2}$ Research Center for Quantum Physics, Huzhou University, Huzhou, 313000, P. R. China\\
$^{3}$ CIC, Instituto Polit\'{e}cnico Nacional, UPALM, CDMX07700, Mexico\\
$^{4}$ Department of Physics, Faculty of Science, Lorestan University, Khorramabad, Iran.\\
$^{5}$ Department of Physics, Faculty of Science, University of Hradec Kr\'{a}lov\'{e}, Rokitansk\'{e}ho 62, 500 03 Hradec Kr\'{a}lov\'{e}, Czech Republic.\\
$^{6}$ Departamento de F\'{\i}sica Te\'orica, At\'omica y Optica and Laboratory for Disruptive\\ Interdisciplinary Science (LaDIS), Universidad de Valladolid, 47011 Valladolid, Spain
\\
}
\date{\today}

\begin{abstract}
\textbf{Abstract:} In this manuscript, we implement the generalized uncertainty principle (GUP) with linear and quadratic moment for Schwarzschild  black hole metric in order to study the influence of quantum effect on the thermodynamics and evaporation of black hole. To this end, we first derive the GUP-modified Hawking temperature of a black hole in the semi-classical framework. Due to the existence of the GUP effect, there is a maximum Hawking temperature. We determine the entropy, heat capacity and Helmholtz free energy
with heuristic analysis that investigates the particle absorbed by black hole. Furthermore, we also verify that these quantities are modified by the GUP, the influence of  quantum effect on the black hole phase transition is  discussed in detail.
Then, we analyze the black hole evaporation process in the mentioned framework and examine the obtained results by graphical methods and compare them with each other.
We likewise explore the behavior of the event horizon radius, photon sphere radius, and shadow silhouette when influenced by the GUP-corrected Schwarzschild black hole (GCSBH) parameters. We intend to establish restrictions for $\alpha$ by utilizing the event horizon telescope (EHT) data for M87* and Sagittarius A* (Sgr A*). Our findings show that Sgr A* provides more robust constraints. As the parameter $\beta$ grows, the range of constraints for $\alpha$ expands.
For Sgr A* one, we find that the shadow radius is close to the observed value at smaller values of $\alpha$.

\end{abstract}
\maketitle

\section{Introduction}
The geometric description of spacetime in general relativity successfully explains the physical phenomena in astrophysics and cosmology, which has made remarkable achievements in black hole physics. In particular, the LIGO Scientific and Virgo collaboration direct detection of the existence of gravitational waves \cite{Abbott1,Abbott2,Abbott3,Abbott4}, and the EHT captured the image of the supermassive black holes M87* and Sgr A* \cite{AkiyamaL12019,AkiyamaL52019,AkiyamaL62019,AkiyamaL122022,AkiyamaL172022,DoScience2019,GravityCollaborationAA2022}. In 1974, Hawking discovered that black holes have thermal radiation \cite{SW1}, this theory not only overcame the contradictions existing in black hole thermodynamics, but also successfully linked gravity theory, quantum mechanics and thermodynamics. The black hole thermodynamics is established by applying general thermodynamics to general relativity to investigate the properties of black holes, the core idea is that the black hole is considered a thermodynamic system, in which the mass of black holes is equal to energy, the entropy is proportional to the area of the event surface of the black hole, and the surface gravity is taken as temperature. The study of the corrected uncertainty principle  has been a very active field of investigation for the  quantum behavior of various black holes. The intrinsic nature of black holes  can  be analyzed more effectively by investigating the absorbed particles by the black holes in  quantum modified spacetime.

The minimal length effect can be effectively predicted by the quantum geometry \cite{5}, the double special relativity \cite{SW2}, non-commutative geometry \cite{SW3}, and other theories of gravity \cite{4,6}. Interestingly, the minimal length can also be obtained by modifying the commutative relation between the coordinate and momentum operators in quantum mechanics, that is, the GUP, which is considered to be one of the most likely results to realize the theory of gravitational quantum mechanics \cite{1,3,GUP15,GUP16,GUP17,GUP18,GUP19,GUP20,GUP21,GUP22,GUP23,GUP24,GUP25,GUP26}. Newly, the GUP effect is widely investigated in black hole physics.  In Ref.\cite{GUP0}, the authors introduced a quantum correction term to correct the Schwarzschild black hole solution, which naturally avoids the existence of singularities,  this finding encouraged research on the quantum effects of black holes. In Ref.\cite{GUP1}, the authors re-discussed the final evolution behavior of black hole evaporation with the GUP effect, and solved the problem of the destruction of cosmic supervision. In Ref.\cite{GUP2,GUP3}, the authors used quantum tunneling radiation method to discuss the thermodynamic phase transition of dark matter to GUP modified XCDM black hole. In Ref.\cite{9}, the authors used the Hamilton-Jacobi method to study the quantum corrections of the Hawking radiation and entropy of the rotating acoustic black hole by the GUP effect and the modified dispersion relation. In Ref.\cite{GUP4,GUP5}, the GUP modified black holes solution under the Topological Defects bumblebee in bumblebee gravity is derived, the effects of GUP on the temperature, entropy, heat capacity and shadow, are investigated. The quasinormal modes of external field disturbances such as scalar field, electromagnetic field and gravitational field disturbances are calculated in terms of  Wentzel-Kramers-Brillouin (WKB) method. Further, we investigated the correction of the GUP effect to the Schwarzschild black hole line element, and used the P\"osch-Teller method to calculate the GUP modified reflection coefficient and transmission coefficient of the black hole scattering radial wave function \cite{GUP6}.  In addition, we can also see certain papers that have a good discussion and an analysis of thermodynamics properties and evaporation of various types of black holes \cite{17,18,19,20,21,22,23,24,26}.

A great number of studies have been, recently, devoted to investigating the Lorentz invariance violation phenomenon at high energy. These results show that, the high energy region close to the Planck energy scale will inevitably cause some physical laws to be corrected. In particular, the double special relativity constructed by the deformation Heisenberg uncertainty principle in quantum mechanics to reflect the modified description of the energy momentum relationship of relativistic particles\cite{29,30,31}.
It is well known that coordinate and momentum operators are the most basic physical quantities in quantum mechanics, the minimum length effect can be reflected by modifying the Heisenberg uncertainty principle. The existence of minimum measurable momentum is also confirmed in the study of black hole thermodynamics, that is, the extended uncertainty principle (EUP)\cite{32,36,37,gup7,gup8}. Based on this, the uncertainty relation is also extended to other algebraic forms such as EGUP, which can simultaneously characterize the existence of minimum length and momentum \cite{40}. Note that the EGUP effect on the thermodynamic properties of Reissner-Nordstr\"om black hole and Bardeen black hole is investigated, the effective limit of quantum effect on the Hawking temperature is analyzed\cite{gup9,gup10}.

Our work is motivated by Ref.\cite{gup11,n3,n4,n5} where thermodynamics and evaporation of the black hole are investigated. To this end, we will consider the higher order forms of GUP with linear and quadratic moments in the presence of new properties and different algebraic structures, which can be expressed as follows\cite{GUP5}:
\begin{eqnarray}
 \Delta x\geq \frac{\hbar }{2{\Delta p}} \left(1-\frac{ \alpha  l_p}{\hbar}\Delta p+\frac{ \beta  l_p^2}{\hbar^2}(\Delta p)^2\right),
\label{s1}
\end{eqnarray}
where $\alpha$ and $\beta$ are dimensionless GUP parameters, $l_p$ denotes the Planck length, and  $l_{p}=\sqrt{\frac{\hbar G}{c^3}}$
. On this basis, we will use the heuristic method \cite{26} to investigate the influence of GUP effect on the heat capacity, entropy and Helmholtz free energy
of Schwarzschild black hole.

An event horizon is a crucial characteristic that describes a black hole; it is the boundary beyond which particles are unable to escape. All physical particles, including light, are trapped inside the event horizon by this strong gravitational field. Light, however, can escape beyond this boundary \cite{SyngeMNRAS131}.
The material surrounding a black hole that is drawn inward is referred to as accretion. The accretion heats up over time as a result of viscous dissipation and releases bright radiation at several frequencies, including radio waves that radio telescopes may detect. The material being accumulated forms a luminous backdrop with a shaded region above it referred to as the black hole shadow \cite{LuminetAA1979}.
While the notion of the black hole shadow \cite{PerlickPR2022,FengEPJC2020,PerlickPRD2015,PerlickPRD2018,VagnozziCQG2023,KocherlakotaPRD2021,WangEPJC2021,CunhaGRG2018} has been present since the 1970s, it was not until Falcke et al. \cite{FalckeApJL13} that the proposal to image the black hole shadow at the center of our Milky Way was initially introduced.
The recent imaging of black hole shadows in the Messier 87 galaxy and Sagittarius A* \cite{AkiyamaL12019,AkiyamaL52019,AkiyamaL62019,AkiyamaL122022,AkiyamaL172022} by the Event Horizon Telescope has sparked widespread interest in current literature.
This phenomenon serves as a focal point for extracting insights into spacetime geometry deviations\cite{LambiaseEPJC2023, AtamurotovPRD2013,AbdujabbarovSS2016,AbdikamalovPRD2019,AtamurotovPRD2015,AtamurotovCPC2023,BelhajPLB2021,BelhajCQG2021,CunhaPLB2017,WeiJCAP2019,LingPRD2021,TsukamotoPRD2018,HeidariPoDU2024,AraujoFilhoCQG2024}, which may arise from various alternative theories of gravity \cite{PantigEPJC2022,PantigAP2023,Uniyal2023,KhodadiPRD2022,KhodadiJCAP2021,PanotopoulosPRD2021} or the astrophysical environment surrounding the black hole \cite{PantigJCAP2022,XuJCAP2018,CapozzielloJCAP2023,Capozziello2023,KonoplyaPLB2019,KonoplyaApJ2022,AnjumPoDU2023}. In this study, our objective is to constrain the GUP parameter $\alpha$ while holding $\beta$ at a fixed value, utilizing an approach extended from Refs. \cite{PerlickPRD2015,PerlickPR2022} that focused on the shadow radius instead of the angular radius. We further investigate the behavior of the black hole shadow using these constraints.

This paper is composed as follows: First, we introduce a correction factor into the event horizon area of Schwarzschild black hole, and derive a new GUP from the intermediate ground of black hole entropy.  Then, we investigate the influence of GUP effect on the heat capacity, entropy and Helmholtz free energy
of black hole, and discuss the thermodynamic phase transition of black hole in detail with graphical methods. Based on the Stefan-Boltzmann radiation law, we also study the GUP effect on the black hole evaporation process. Next, the radius of the BH shadow is determined and compared to observations of M87* and Sgr A* , which allows us to constrain the GUP parameter. Finally, we summarize a short conclusion.

\section{Black holes thermodynamics: a heuristic analysis}
\label{sec:Black holes thermodynamics: a heuristic analysis}
At present, the uncertainty principles of different parameter symbols are obtained by modifying the Heisenberg uncertainty principle, including the GUP with minimal length effect, the EUP with the minimal momentum, and other correction forms, which is widely used in black hole physics. In order to get a new form of GUP algebra, the correction of thermodynamic quantities is reasonable and correct, we naturally introduce a correction factor in the horizon area of Schwarzschild black hole, so as to indirectly correct the existing linear and quadratic moments GUP and derive a new generalized uncertainty principle.

In the semiclassical framework, the relationship between Hawking temperature and surface gravity in a stationary and spherically symmetric black hole is as follows:
	\begin{eqnarray}
	{T_0} = \frac{{\kappa \hbar}}{{4\pi }},\label{s2}
	\end{eqnarray}
where $\kappa$ represents the surface gravity,  and the reduced plank constant $\hbar$  is the quantum effect of black hole thermal radiation. Since the modified thermodynamic properties are equivalent to the minimum length effect, the Hawking temperature in Eq. \eqref{s2} in the semiclassical framework  should be modified by the GUP quantum effect. As well as, the Bekenstein-Hawking black hole entropy is given by
 \begin{eqnarray}
	{S_{0}} = \frac{A}{{2\hbar }},\label{s02}
\end{eqnarray}
where $A$ represents the area of the horizon, Black hole thermodynamics is very similar to ordinary thermodynamics, which can be written as \cite{26}
\begin{eqnarray}
dM = \frac{\kappa}{{8\pi}}dA+\sum_{i}X_{i}dx_{i},\label{s012}
\end{eqnarray}
where the phrase $x_{i}$ describes the thermodynamic quantities of the black hole, such as charge and angular momentum, while the $X_{i}$ term is a generalized force, including the angular velocity and electrostatic potential of the black hole. Based on this, the temperature is given by

\begin{eqnarray}
{T} = \bigg(\frac{\partial M}{\partial S}\bigg)_{x_{i}}=
\frac{dA}{dS}\times\bigg(\frac{\partial M}{\partial A}\bigg)_{x_{i}}=
\frac{dA}{dS}\times\frac{\kappa}{{8\pi}}. \label{ga1}
\label{s002}
\end{eqnarray}
It can be seen from the Eq. \eqref{ga1} that Hawking temperature depends on the horizon area, entropy and surface gravity. In order to indirectly find a new GUP algebraic form in the black hole entropy, we consider the case where a particle disappears after being captured by the black hole, and the information is then transmitted to observers outside the black hole horizon. In addition, the smallest
increase of the black hole horizon area $\Delta A $ is written as

\begin{eqnarray}
	\Delta A \sim \sigma m, \label{s3}
\end{eqnarray}
where $m$ is the mass of the trapped particles, and $\sigma$ means the size of the particles. In this case, the change in entropy of the black hole is written as
\begin{eqnarray}
{\Delta S} \simeq \frac{dS}{dA}\Delta A.
\label{s03}
\end{eqnarray}
According to information theory,  we can consider ${\left( {\Delta S} \right)_{\min }} = \rm \ln 2$
and $(\Delta A)_{min}=0$ for a classical particle (a point-like object) \cite{26}. Based on this, there is no precise orbit of microscopic particle in quantum mechanics, and the state is described by wave packet. On the one hand, the width of the wave packet can represent the position uncertainty of the particle $\Delta x$. On the other hand, in the process of measuring the motion of particle, the particle mass must be greater than the momentum uncertainty $\Delta p$. Otherwise, the relativistic effect leads to the generation of particle pairs, which makes the measurement meaningless.
By considering equation \eqref{s03} horizon area is rewritten as
\begin{eqnarray}
	A \sim \sigma m \geqslant \Delta x\Delta p,\label{s04}
\end{eqnarray}
it can be found that the horizon area is not less than the product of coordinate and momentum uncertainty, and the GUP effect will correct the result of Beckenstein-Hawking black hole entropy. By further substituting Eq. \eqref{s2} into Eq. \eqref{s04}, a new GUP is obtained as follows:
\begin{eqnarray}
\Delta A\geq \gamma_{1}\frac{\hbar }{2} \left(1-\frac{ \alpha  l_p}{\hbar}\Delta p+\frac{ \beta  l_p^2}{\hbar^2}(\Delta p)^2\right),
\label{s06}
\end{eqnarray}
here $\gamma_{1}$ is the correction factor. It can be seen that the minimal increase of the black hole horizon area $(\Delta A)_{min}$  depends on the minimal momentum uncertainty $\Delta p$ . When the minimal momentum uncertainty tends to zero, the minimum horizon area tends to $\frac{\hbar\gamma_{1}}{2}$. In addition, the uncertainty relation in quantum mechanics shows that $\Delta x$ tends to infinity when $\Delta p$ is close to zero. However, $\Delta p$ cannot be arbitrarily small, and the position uncertainty in the corresponding region is limited, which provides some inspiration for finding new GUP from the black hole entropy. In order not to lose generality, we will use these units ($G= c =K_{B}=\hbar =l_{p} = 1$) in the following sections.

\section{Thermodynamics of the Schwarzschild Black Hole}
\label{sec:Thermodynamics of the Schwarzschild Black Hole}
Now, we will investigate the thermodynamics and evaporation time of black holes under the new GUP framework. In the previous section, if a particle is captured  by the Schwarzschild black hole, its position uncertainty cannot be greater than the twice radius of the horizon. By considering the GUP effect,  it can be found that
\begin{eqnarray}
2 r_0\geq \Delta x\geq \frac{1}{2 \Delta p}
  \left(1-\alpha \Delta p+\beta(\Delta p)^2\right). \label{s5}
\end{eqnarray}
Based on this, the effective interval of momentum uncertainty correction by GUP effect is as follows:
\begin{equation}
\frac{\alpha -\sqrt{\left(-\alpha -4 r_0\right){}^2-4 \beta }+4 r_0}{2 \beta }\leq \text{$\Delta $p}\leq \frac{\alpha +\sqrt{\left(-\alpha -4 r_0\right){}^2-4 \beta }+4 r_0}{2 \beta }. \label{s006}
\end{equation}
In this case, by the substitution Eqs. \eqref{s5} and \eqref{s006} in Eq. \eqref{s06}, we can observe that the horizon area is not less than the product of the correction factor and the effective Planck constant
\begin{eqnarray}
	A \geqslant \gamma_{1} \hbar' \label{s7}
\end{eqnarray}
with
\begin{eqnarray}
	\hbar' =\frac{1}{2} \left(1-\frac{\alpha  \left(\alpha -\sqrt{\left(-\alpha -4 r_0\right){}^2-4 \beta }+4 r_0\right)}{2 \beta }+\frac{\left(\alpha -\sqrt{\left(-\alpha -4 r_0\right){}^2-4 \beta }+4 r_0\right){}^2}{4 \beta }\right).
	\label{s8}
\end{eqnarray}
If the GUP parameters $\alpha$ and $\beta$ tend to zero, the GUP degenerates into the uncertainty principle, and the effective Planck becomes the reduced Planck constant.We consider these findings, the GUP effect on the Hawking temperature of the Schwarzschild black hole reads
\small{
\begin{eqnarray}
	\label{g2}
T_{\rm GUP}&=& \frac{{\kappa \hbar'}}{{2\pi }}
\nonumber\\
&=&\frac{\kappa}{4 \pi } \left(1+\frac{\left(\alpha -\sqrt{\left(-\alpha -4 r_0\right){}^2-4 \beta }+4 r_0\right){}^2}{4 \beta }-\frac{\alpha  \left(\alpha -\sqrt{\left(-\alpha -4 r_0\right){}^2-4 \beta }+4 r_0\right)}{2 \beta }\right).\label{s9}
\end{eqnarray}}
 Further, we consider the GUP parameters as small quantity, and the Hawking temperature is expanded by Taylor series
\small
{\begin{eqnarray}
\label{g2}
T_{\rm GUP}&=&\frac{\kappa }{4 \pi }\bigg[1-\frac{\alpha  }{2 \beta } \left(\alpha -4 r_0 \sqrt{\left(\frac{\alpha }{4 r_0}+1\right){}^2-\frac{4 \beta }{16 r_0^2}}+4 r_0\right)
+\frac{1}{4 \beta } \left(\alpha -4 r_0 \sqrt{\left(\frac{\alpha }{4 r_0}+1\right){}^2-\frac{4 \beta }{16 r_0^2}}+4 r_0\right){}^2\bigg]
\nonumber\\
&=&
\frac{\kappa  }{4 \pi }\bigg[1-\frac{\alpha  }{2 \beta }\left(\alpha -4 r_0 \left(\frac{\alpha }{4 r_0}-\frac{2 \beta }{16 r_0^2}+1\right)+4 r_0\right)+\frac{1}{4 \beta }\left(\alpha -4 r_0 \left(\frac{\alpha }{4 r_0}-\frac{2 \beta }{16 r_0^2}+1\right)+4 r_0\right){}^2\bigg].
\end{eqnarray}}
Finally, the Eq. \eqref{s9} can be rewritten as follows
\begin{eqnarray}
T_{\rm GUP}=\frac{\kappa }{4 \pi } \left(1-\frac{\alpha }{4 r_0}+\frac{\beta }{16 r_0^2}\right).\label{g1}
\end{eqnarray}
If the GUP parameters $\alpha$ and $\beta$ tend to zero, the modified temperature reduces to the stationary black hole Hawking temperature which is the same as Eq. \eqref{s2}. In the Schwarzschild black hole, we consider that the horizon radius is equal to twice the mass for values of $c=G=1$ and $\kappa=\frac{1}{2 r_{0}}$ \cite{26}. Therefore, the Eq. \eqref{g1} versus $M$ can be considered as
\begin{eqnarray} \label{g2}
T_{\rm GUP} =\frac{1}{16 \pi  M}\left(1-\frac{\alpha }{8 M}+\frac{\beta }{64 M^2}\right),
\end{eqnarray}
also we have
\begin{eqnarray}\label{rh}
r_{\rm GUP}=2 M_{\rm GUP},
\end{eqnarray}
with
\begin{eqnarray}\label{g02}
M_{\rm GUP} =M \left(1+\frac{\alpha }{8 M}-\frac{\beta }{64 M^2}\right).
\end{eqnarray}

\begin{figure}[H]
	\centering
	\includegraphics[width=.45\textwidth]{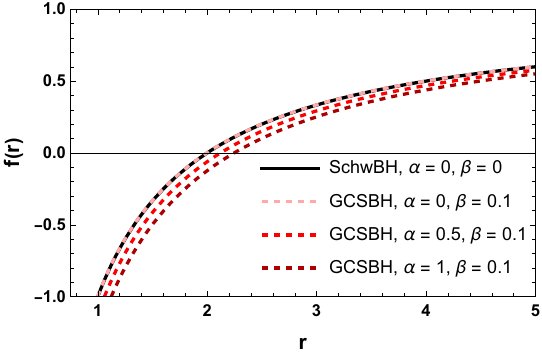}
	\caption{\label{fig1}
		Plotting the lapse function $f(r)$ for various $\alpha$ values, setting $M = 1$.}
\end{figure}

\begin{figure}[h]
	\centering
		\includegraphics[width=.45\textwidth]{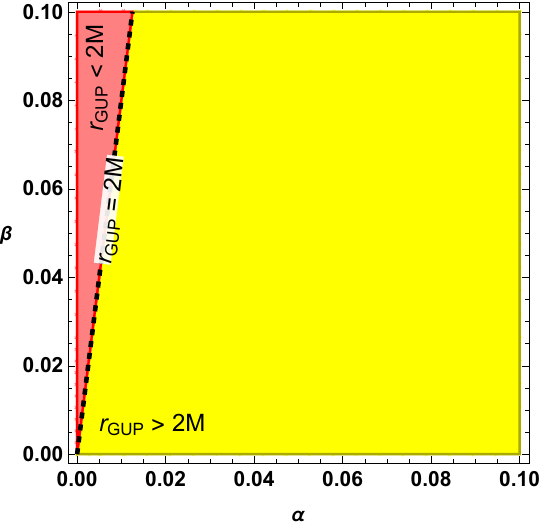}
	\caption{\label{fig2} Showing the connection between $r_{\text{GUP}}$ and $r_{\text{H}}$ in terms of  the parameter space ($\alpha$, $\beta$), setting $M = 1$.}
\end{figure}

The following line element describes the static and spherically symmetric metric for the Schwarzschild black hole with quantum corrections imposed by the GUP (the GUP-corrected Schwarzschild black hole (GCSBH) solution), which is produced by substituting the mass $M$ with the mass GUP, $M_{\rm GUP}$, (or substituting $r_{\rm GUP}$ for the Schwarzschild horizon radius, $r_{\rm H} = 2M$)
\begin{subequations}\label{Metric}
\begin{equation}\label{Metric1}
	ds^{2} = -f(r)dt^{2}+f(r)^{-1}dr^{2}+r^{2} \left(d\theta^{2}+\sin^{2}\theta d\varphi^{2}\right).
\end{equation}
with
\begin{equation}\label{MetricFunction1}
f(r) = 1-\frac{2M_{\rm GUP}}{r}.
\end{equation}
\end{subequations}
The metric \eqref{Metric1} resembles the Schwarzschild form, with the modification factor $(\frac{\alpha }{8 }-\frac{\beta }{64 M})$ being independent of coordinates. This solution represents a vacuum scenario and reveals that the GCSBH metric characterizes a black hole with a single horizon.
 Despite the singularity not being eliminated, we will demonstrate that it cannot be reached \cite{CarrJHEP2015}.
The GCSBH solution \eqref{Metric1} encompasses spherical Schwarzschild spacetimes in the limit as $\alpha$ and $\beta$ approach $0$. Moreover, in the scenario where $M$ tends to $0$, and both $\alpha$ and $\beta$ approach $0$, Eq. \eqref{Metric1}  describes a flat spacetime.
Indeed, as $r$ approaches infinity, the lapse function $f(r)$ converges to the Schwarzschild one. Closer to the black hole, as depicted in Fig. \ref{fig1}, the influence of the GCSBH parameter $\alpha$ becomes more pronounced compared to the GCSBH parameter $\beta$.
Setting $f(r)=0$ resolves the coordinate singularity in Eq. \eqref{Metric} and allows for the definition of the horizons of the GCSBH as previously discussed. Depending on the values of $\alpha$ and $\beta$, the equation $f(r) = 0$ can have a positive real root.
In terms of the parameter space ($\alpha$, $\beta$), the relationship between $r_{\text{GUP}}$ and $r_{\text{H}}$ can be categorized into three mass regimes: $r_{\text{GUP}} > 2M$, $r_{\text{GUP}} = 2M$, and $r_{\text{GUP}} < 2M$, as illustrated in Fig. \ref{fig2}. We have identified two regions:
I)Yellow Region: In this region, the single horizon resulting from $f(r) = 0$ is greater than $2M$.
II)Red Region: Conversely, in this region, the single horizon arising from $f(r) = 0$ is less than $2M$.
The boundaries indicated by black dashed lines separate these regions, corresponding to the expression $r_{\text{GUP}} = 2M$ derived from $f(r) = 0$, signifying the situation where the choice of parameter space ($\alpha$, $\beta$) leads $r_{\text{GUP}}$ to match the standard Schwarzschild radius.
Besides, the significance of the GCSBH parameter $\alpha$ is evident when compared to the GCSBH parameter $\beta$ in delineating the behavior of the GCSBH horizon relative to the standard Schwarzschild radius.

\begin{figure}[htbp]
	\centering
	\includegraphics[height=5cm,width=16cm]{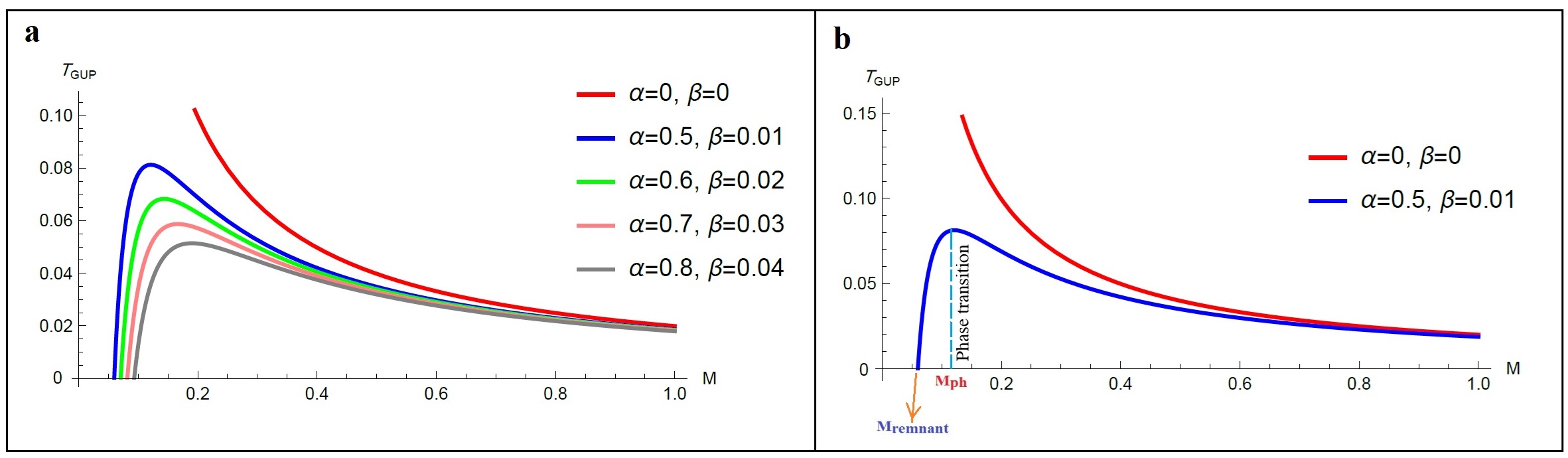}
	\caption{\label{fig3}The GUP-corrected temperature versus the mass of the Schwarzschild black hole for different values of $\alpha$ and $\beta$ (in FIG. (a))
		and for a special value of$\alpha$ and $\beta$ by specifying the amount of $M_{\rm ph}$ and $M_{\rm remnant}$ (in FIG. (b)). }
\end{figure}

In order to have a more detailed analysis and provide a better understanding, we plot the thermal functions. We first discuss the influence of GUP effect on the Hawking temperature. In Fig. \ref{fig3}, we depict the Schwarzschild black hole temperature function via the mass for different values of the GUP parameters, it can be observed that Hawking temperature increases first and then decreases with the mass. For the larger GUP parameter $\beta$, the Hawking temperature has a smaller value at the same mass.
The minimal non-zero mass and the remnant mass have been  specified in Fig. \ref{fig3}, and we can see they depend clearly on the amount of GUP parameters $\alpha$ and $\beta$. Note that due to the existence of the GUP effect, there is a phase transition at the maximum Hawking temperature, which entails that its corresponding heat capacity has a value of zero at this point.This interesting phenomenon will not appear in the uncorrected Schwarzschild black hole.
From the above relation and by considering $T_{\rm GUP} =0$, we can remnant mass versus $\alpha$ and $\beta$ as
\begin{eqnarray}\label{MassRem}
M_{\text{remnant}}=\frac{1}{16} \left(\sqrt{\alpha ^2-4 \beta }+\alpha \right).\label{g3}
\end{eqnarray}
Equation \eqref{MassRem} gives as a constraint for $\alpha$ and $\beta$ and according it $\alpha\ge 2\sqrt\beta$.\\
In Table \ref{remnantMass} we reported the remnant mass for different $\alpha$ and $\beta$. According to the data reported in Table \ref{remnantMass} for the same $\alpha$, when $\beta$ increases the remnant mass decreases and for the same $\beta$, when $\alpha$ increases the remnant mass increases.
\\
\begin{table}[h!]
	\caption{The remnant mass of the black hole for different amounts of $\alpha$ and $\beta$.}\label{remnantMass}	\centering
	\begin{tabular}{c c c}
		\hline
		$\alpha$ &
		$\beta$ & $M_{\rm remnant}$ \\
		\hline
	0.5 \,\,\,& 0.01 \,\,\,& 0.05989 \\
	0.5 \,\,\,& 0.03 \,\,\,& 0.05378 \\
	0.5 \,\,\,& 0.05 \,\,\,& 0.04522 \\
	0.3 \,\,\,& 0.02 \,\,\,& 0.02500 \\
	0.5 \,\,\,& 0.02 \,\,\,& 0.05702\\
	0.7 \,\,\,& 0.02 \,\,\,& 0.08377 \\
	\hline
	\end{tabular}
\end{table}

Also, by putting zero value for derivative of the temperature in Eq. \eqref{g2}, we can calculate the mass associated with  phase transition point of the system
\begin{eqnarray}
M_{\text{ph}}=\frac{1}{8} \left(\sqrt{\alpha ^2-3 \beta }+\alpha \right).\label{g4}
\end{eqnarray}

Next, we will use the analytical method to study the influence of GUP effect on the black hole entropy function. It is well known that the GUP correction entropy of the Schwarzschild black hole is related to the effective Planck and the horizon area, as shown below
\begin{eqnarray}
S_{\rm GUP} = \int {\frac{{dS}}{{dA}}} dA \simeq \int {\frac{{{{\left( {\Delta S} \right)}_{\min }}}}{{{{\left( {\Delta A} \right)}_{\min }}}}} dA \simeq \int {\frac{{dA}}{{4\hbar '}}}\frac{\gamma_{1}}{\rm \ln 2}.\label{s12}
\end{eqnarray}
and
\begin{eqnarray}
\frac{{dA}}{{dS}} = \frac{{{{\left( {\Delta A} \right)}_{\min }}}}{{{{\left( {\Delta S} \right)}_{\min }}}} = 4\hbar'\frac{\rm \ln 2}{\gamma_{1}}.\label{s13}
\end{eqnarray}
In this case, we consider the modification of black hole entropy by the new GUP effect, and obtain
\begin{equation}\label{s14}
\begin{split}
S_{\rm GUP}=&\frac{\gamma_{1}}{4\rm \ln 2}\left[2 \pi \alpha r_0+4 \pi r_0^2+\frac{1}{4} \pi\left(\alpha+4 r_0\right) \sqrt{\alpha^2-4 \beta+8 \alpha r_0+16 r_0^2}\right.\\
&\left.-\pi \beta L n\left(\alpha+\sqrt{\alpha^2-4 \beta+8 \alpha r_0+16 r_0^2}+4 r_0\right)\right].
\end{split}
\end{equation}
We observe the influence of the GUP effect on the entropy function of the black hole from Eq. \eqref{s14}. Interestingly, due to the existence of the  quadratic moment  GUP effect ( $\beta\neq0$ ), a logarithmic term appears in the Schwarzschild black hole, which is similar to the results of other quantum correction entropy functions. Further, the GUP parameters as small quantities, and it can be found that
\begin{eqnarray}
S_{\rm GUP}=\frac{\gamma_{1}}{4\rm \ln 2} \pi  \left(8 r_0^2-\frac{\beta }{2}-\frac{3 \alpha  \beta }{8 r_0}+4 \alpha  r_0-\beta  \ln \left(8 r_0\right)\right). \label{g5}
\end{eqnarray}
If we ignore the existence of GUP effect $(\alpha=\beta=0)$, the result is consistent with the Schwarzschild black hole entropy in the semiclassical framework \cite{26}.
It should be reproduce the standard result of the entropy $S = \pi r_{0}^{2}$ when $\alpha = \beta = 0$. This requires that the calibration factor yield $2 \gamma_{1} =\rm ln\, 2$ \cite{26}.

\begin{figure}[htbp]
	\centering
	\includegraphics[height=8cm,width=10cm]{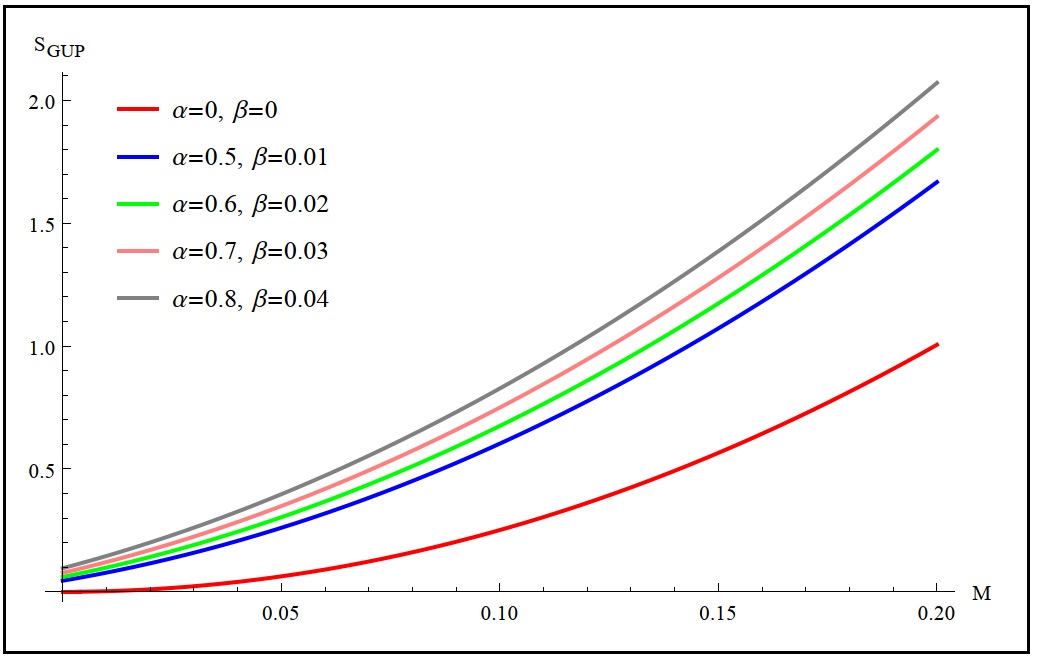}
	\caption{\label{fig4}The GUP-corrected entropy versus the mass of the Schwarzschild black hole for various $\alpha$ and $\beta$.}
\end{figure}
In Fig. \ref{fig4}, we plotted the change of entropy function with mass when the GUP parameter was assigned different values,this can be observed that the GUP modified black hole entropy increases with the increase of mass.

Subsequently, in order to  analyze the correction of the quantum effect on the black hole phase transition, it is generally known that the heat capacity is zero corresponding to the black hole phase transition point. Therefore, we will use the analytical method to calculate the heat capacity. According to the entropy and temperature expressed by the black hole have been obtained, the heat capacity expression reads
\begin{eqnarray}
C_{\text{GUP}}=T\frac{\partial S}{\partial T}=\frac{T \frac{\partial S}{\partial M}}{\frac{\partial T}{\partial M}},\label{s11}
\end{eqnarray}
which leads to
\begin{eqnarray}
C_{\text{GUP}}=-\frac{16 \pi  M^3 \left(\frac{\beta }{64 M^2}-\frac{\alpha }{8 M}+1\right) \left(8 \alpha +\frac{3 \alpha  \beta }{16 M^2}-\frac{\beta }{M}+64 M\right)}{3 \beta +64 M^2-16 \alpha  M},\label{s11}
\end{eqnarray}	
we can clearly find that if the GUP parameters are both zero, that is, the heat capacity without quantum correction is $	C_{\text{GUP}}=-16 \pi  M^2$, the result is completely consistent with the heat capacity of the semiclassical Schwarzschild black hole \cite{gup11}, which also verifies its correctness.
\begin{figure}[htbp]
	\centering
	\includegraphics[height=5cm,width=16cm]{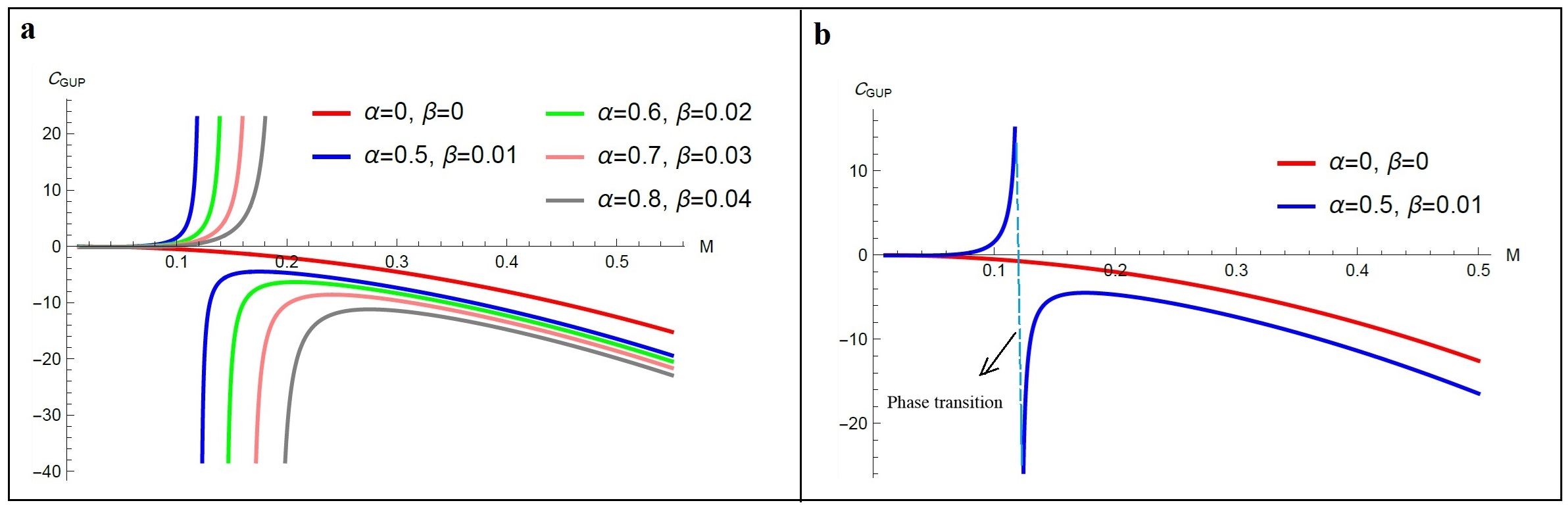}
	\caption{\label{fig5}The GUP-corrected specific heat versus the mass of the Schwarzschild black hole for different values of $\alpha$ and $\beta$ (in Fig. (a))
		and for a special value of $\alpha$ and $\beta$ by specifying the phase transition (in Fig. (b)).}
\end{figure}
In Fig. \ref{fig5}, the heat capacity with the change of mass is plotted under the GUP correction. As expected, the GUP effect makes the heat capacity appear positive. If the GUP parameters value is fixed, we find that with the increase of black hole mass, the heat capacity function first increases rapidly and then decreases slowly. It should be emphasized that the heat capacity is closely related to the stability of the black hole. In more detail, the region with positive heat capacity has local stability. On the contrary, if the heat capacity is negative, the black hole will be unstable. Furthermore, if we consider the heat capacity is zero, the black hole mass will not change with the change of temperature, that is, the remnant mass $M_{\rm remnant}$ is obtained. In addition, because the Hawking temperature and the specific heat function have discontinuous regions, this means that the GUP effect causes a phase transition phenomenon in the black hole in Fig. \ref{fig3} and Fig. \ref{fig4}. In order to better understand the influence of GUP effect on the phase transition of Schwarzschild black hole, we mark the phase transition mass point as $M_{\rm Ph}$  in Fig. \ref{fig3}. If the mass value is less than $M_{Ph}$, the black hole has a stable behavior, while instability will occur in other regions.

Up to now, we have given the expressions of Hawking temperature, heat capacity and entropy of the GUP modified Schwarz black hole. Based on this, we can further derive the GUP-corrected Helmholtz free energy function from the usually famous analytical formalism \cite{GUP12,GUP13,GUP14}
\begin{equation}\label{sq1}
\begin{aligned}
\mathcal{F}_{\rm GUP}&=M_{\rm GUP}-T_{\rm GUP}S_{\rm GUP} \\
&=\frac{M}{2}+\frac{1}{64 M}\bigg[4 \alpha  M-\frac{\beta }{2}-\left(\frac{\beta }{64 M^2}-\frac{\alpha }{8 M}+1\right) \left(-\frac{\beta }{2}-\frac{3 \alpha  \beta }{16 M}+8 \alpha  M-\beta  \log (16 M)\right)\bigg],
\end{aligned}
\end{equation}
If we ignore the influence of GUP effect on the Helmholtz free energy in Eq. \eqref{sq1}, the result degenerates to obtained result in Ref. \cite{gup11}.
\begin{figure}[htbp]
	\centering
	\includegraphics[height=5cm,width=16cm]{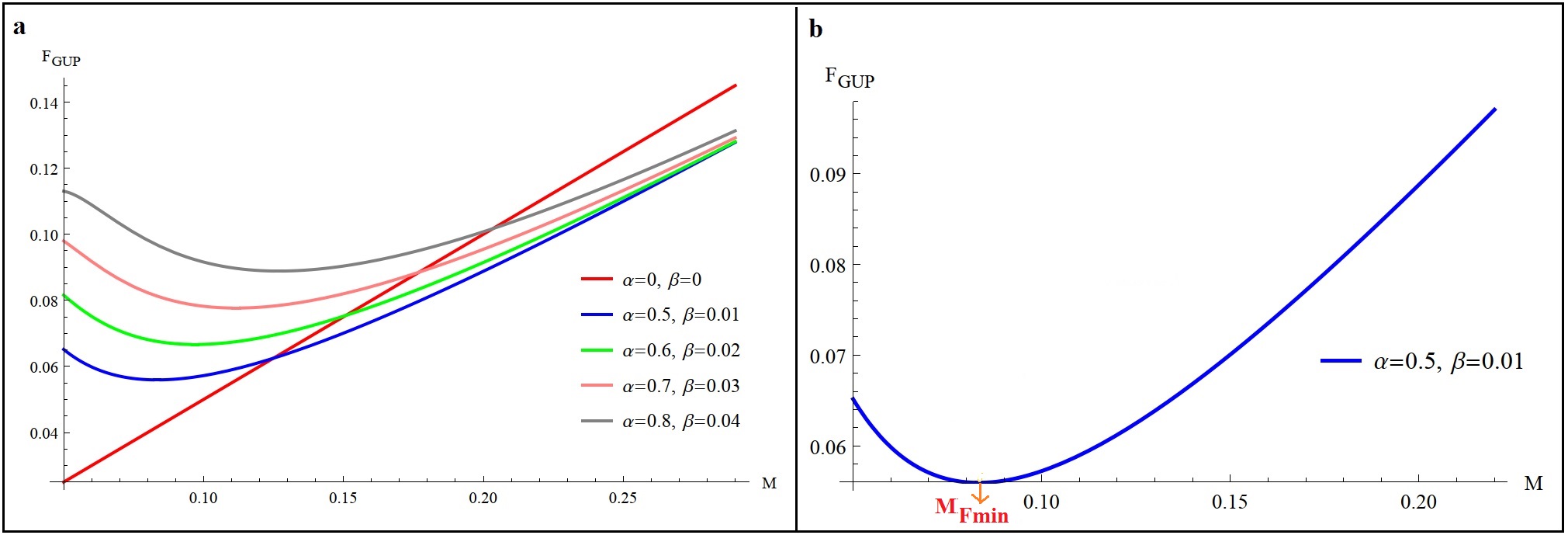}
	\caption{\label{fig6}The GUP-corrected Helmholtz free energy versus the mass of the Schwarzschild black hole for different values of $\alpha$ and $\beta$ (in FIG. (a))
		and for a special  value of$\alpha$ and $\beta$ by specifying the amount of $M_{Fmin}$ (in FIG. (b)).}
\end{figure}
In Fig. \ref{fig6}, when the GUP parameter takes different values, the relationship between the Helmholtz free energy and the mass is drawn. We can observe that the GUP effect causes the Helmholtz free energy to have a minimal $\mathcal{F}_{\rm GUP}$ in the non-zero mass region. On the one hand, For the same mass, the Helmholtz free energy increases with the increase of GUP parameters. On the other hand,
with the increase of mass, Helmholtz free energy decreases first and then increases.

\section{Energy emission rate}
In this section,  the behaviour of the energy emission rate versus frequency has been studied.
It should be noted that the quantum fluctuations leads to  certain pairs of particles near the black hole horizons  and the particles with the positive energy can escape through tunneling from the black hole and in this case the Hawking radiation occurs .
This phenomenon called the Hawking radiation. The Hawking radiation play a crucial role in  the evaporation of the black holes in a certain time.
	
\begin{figure}[htbp]
	\centering
	\includegraphics[height=7cm,width=10cm]{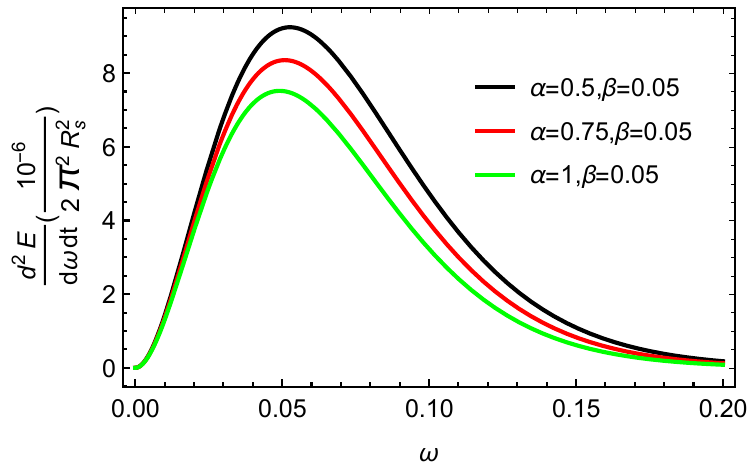}
	\caption{\label{fig7}The energy emission rate versus $\omega$ for different $\alpha$ and $\beta=0.05$.}
\end{figure}

In most instances,  the remnant mass is considered as the final mass of the black hole and this causes to the final mass of the black hole is not zero. Importance of the existence of the remnant mass is determined by the fact that all information of the black hole not to be missed.
we can cosider the energy emission rate as the following form 
	\begin{equation}
		\frac{d^2E}{d\omega dt}=\frac{2\pi^2 \sigma_{lim}\omega^3}{e^{\frac{\omega}{T}}-1},
	\end{equation}
where   $T$ is the Hawking temperature and $\sigma_{lim}=\pi R_{\rm sh}^2$ where $R_{\rm sh}$ is the radius of the shadow, for the outer event horizon that it obtained according to Eq. \eqref{g2} and $\omega$ represents the frequency of the photon.
In what follows,  the corresponding energy emission rate has been  explored.
In Fig. \ref{fig7} we plotted the energy emission rate versus the frequency $\omega$ for different values of $\alpha$ and $\beta$. It is obvious that to know the maximum of the energy emission rate we should drevate respect to $\omega$. According to Fig. \ref{fig7}, we see that for the same $\beta$, when $\alpha$ increases the peak of the energy emission rate increases and also it shift to low frequency.

Also according to the energy emission rate equation we see that when frequencies goes to zero as well as to infinity the emission rate goes to zero.

\section{Evaporation Time}
By considering  the Stefan-Boltzmann radiation law and evaporating the black hole under GUP, we can calculate  the black hole's   lifetime   as
 \begin{eqnarray}
\frac{dM_{\text{GUP}}}{dt}= -\tilde{\alpha } \tilde{a}\sigma  T_{\text{GUP}}^4,  \label{s22}
\end{eqnarray}
here, $\tilde{\alpha }$ means the grey factor,  $\tilde{a}$ is the radiation constant, and $\sigma$ represents the cross-section area \cite{n3,n4,n5}. By considering the photon capture cross section as $\sigma = 27 \pi M^2$  in the
geometric optics approximation and  $T=\frac{\kappa }{4 \pi }$ or $M=\frac{1}{16 \pi  T}$, we can write
 \begin{eqnarray}
\frac{dM_{\text{GUP}}}{dt}=-\frac{ 27 \pi  \tilde{\alpha } \tilde{a}}{(16 \pi )^2}T_{\text{GUP}}^2,  \label{s23}
\end{eqnarray}
In this framework for our GUP black hole, the mass evolution form becomes
 \begin{eqnarray}
\frac{dM_{\text{GUP}}}{dt}=-\frac{\eta  }{M^2}\left(1-\frac{\alpha }{4 M}+\frac{\beta }{32 M^2}\right),  \label{s24}
\end{eqnarray}
where
\begin{eqnarray}
\eta =\frac{27 \pi  \tilde{\alpha } \tilde{a}}{(16 \pi )^4}.  \label{s25}
\end{eqnarray}
Therefore, the estimate of the time of the evaporation process can be obtained as
\begin{eqnarray}
-\int_{M_i}^{\tilde{M}} M^2 \left(-\frac{\beta }{64 M^2}+\frac{\alpha }{4 M}+1\right) \, dM=\int_0^t \eta  \, dt,  \label{s26}
\end{eqnarray}
where $M_{i}$ is the initial mass of the black hole after evaporation time $t$, and $\tilde{M}$ is the current mass. Based on this, we can get the expression of evaporation time
\begin{eqnarray}
t=\frac{1}{\eta }\bigg(\frac{M_i^3}{3}+\frac{\alpha  M_i^2}{8}-\frac{\beta  M_i}{64}-\frac{\alpha  \tilde{M}^2}{8}+\frac{\beta  \tilde{M}}{64}-\frac{\tilde{M}^3}{3}\bigg). \label{s27}
\end{eqnarray}

In above equation, we can find that the evaporation time depends on the GUP parameters and the black hole mass. From time $t = 0$, the black hole mass $\tilde{M}$ will be smaller than the initial mass $M_{i}$. If we consider the final state of the process, the black hole mass is equal to $M_{\text{remnant}}$, and the lifetime of the black hole is given by

\begin{eqnarray}
t_f=\frac{M_i}{192 \eta } \bigg(64 M_i^2 +24 \alpha  M_i-3 \beta\bigg). \label{s28}
\end{eqnarray}

Furthermore, the time evaporation can be calculated in the absence of the GUP effect as $(\lim t_f=\frac{M_i^3}{3 \eta })$ for the values of $\alpha\rightarrow 0$ and $\beta\rightarrow 0$.
\begin{table}[h]
	\centering
	\caption{The lifetime of the black hole for different amounts of $\alpha$ and $\beta$.}\label{tab2}
	\begin{tabular}{c c c}
		\hline
		$\alpha$ &
		$\beta$ & lifetime \\
		\hline
	0 \,\,\,& 0 \,\,\,& 0.333333 \\
0.5 \,\,\,& 0.01 \,\,\,& 0.395677 \\
0.6 \,\,\,& 0.02 \,\,\,& 0.408021 \\
0.7 \,\,\,& 0.03 \,\,\,& 0.420365 \\
0.8 \,\,\,& 0.04 \,\,\,& 0.432708 \\
\hline
	\end{tabular}
\end{table}
In Table \ref{tab2} we have reported the evaporation time for different $\alpha$ and $\beta$. In fact, different $\alpha$ and $\beta$ belong to different remnant masses.  Therefore when a black hole evaporates, the mass of the black hole radiuses from $M_i$ to $M_f=M_{\rm remnant}$, and the interval time is named as lifetime. we see that when the remnant mass increases the lifetime increases.

\section{GUP-corrected Schwarzschild Black hole shadow and constraints}
In this section, we initially investigate the shadow radius of the GCSBH solution, which is adjusted by incorporating the parameters $\alpha$ and $\beta$ from the GUP approach.
The boundary of the black hole shadow, observable to a distant observer, defines the visible region of the photon region by distinguishing between capture and scattering orbits.
The photon area essentially marks the edge of the spacetime region, which in the case of spherically symmetric spacetime, corresponds to the photon sphere.
Subsequently, we employ empirical data for M87* and Sgr A* from the EHT collaboration, as detailed in Table \ref{tab5}, to constrain the $\alpha$ parameter.
It is possible to derive the photon sphere radius and critical impact parameter for the conventional Schwarzschild metric as $3M$ and $3\sqrt{3}M$, respectively. In the present scenario, with regard to Eq. \eqref{g02}, these radii could be rearranged as follows
\begin{equation}\label{rphbcRadii}
r_{\rm ph} = 3M_{\rm GUP} \equiv 3 M \left(1+\frac{\alpha }{8 M}-\frac{\beta }{64 M^2}\right), \quad b_{\rm c} = 3\sqrt{3} \,M_{\rm GUP} \equiv 3\sqrt{3}\, M \left(1+\frac{\alpha }{8 M}-\frac{\beta }{64 M^2}\right).
\end{equation}
The numerical findings for the GCSBH event horizon radius $r_{\rm GUP}$, the photon sphere radius $r_{\rm ph}$, and the critical impact parameter $b_{\rm c}$ for various values of the GCSBH parameters can be found in Tables \ref{tab3} and \ref{tab4}, according to Eqs. \eqref{rh}, \eqref{g02}, and \eqref{rphbcRadii}.
Increasing $\alpha$ and $\beta$ leads to an increase and decrease, respectively, in $r_{\rm GUP}$, $r_{\rm ph}$, and $b_{\rm c}$ compared to a Schwarzschild black hole, where $\alpha = 0 =\beta$.
\begin{table}[h!]
	\caption{Values of the GCSBH event horizon radius $r_{\rm GUP}$, photon sphere radius $r_{\rm ph}$, and critical impact parameter $b_{\rm c}$ for various $\alpha$ values, where $\beta = 0.1$ and $M=1$.}
	\label{tab3}\centering
	\begin{tabular}{ccccccccccccccc}
		\hline
		{$\alpha$} &\,\,\,\,\,&{$0$}&\,\,\,&{$0.1$} &\,\,\,& {$0.3$}&\,\,\,&{$0.5$} &\,\,\,& {$0.7$} &\,\,\,& {$0.9$} &\,\,\,& {$1$}\\
		\hline
		{$r_{\rm GUP}$} &\,\,\,\,\,&{$1.99688$}&\,\,\,&{$2.02188$} &\,\,\,& {$2.07188$}&\,\,\,&{$2.12188$} &\,\,\,& {$2.17188$} &\,\,\,& {$2.22188$} &\,\,\,& {$2.24688$}\\
		{$r_{\rm ph}$} &\,\,\,\,\,&{$2.99531$}&\,\,\,&{$3.03281$} &\,\,\,& {$3.10781$}&\,\,\,&{$3.18281$} &\,\,\,& {$3.25781$} &\,\,\,& {$3.33281$} &\,\,\,& {$3.37031$}\\
		{$b_{\rm c}$} &\,\,\,\,\,&{$5.18803$}&\,\,\,&{$5.25299$} &\,\,\,& {$5.38289$}&\,\,\,&{$5.51279$} &\,\,\,& {$5.6427$} &\,\,\,& {$5.7726$} &\,\,\,& {$5.83755$}\\
		\hline
	\end{tabular}
\end{table}
\begin{table}[h!]
	\caption{Values of the GCSBH event horizon radius $r_{\rm GUP}$, photon sphere radius $r_{\rm ph}$, and critical impact parameter $b_{\rm c}$ for various $\beta$ values, where $\alpha = 0.1$ and $M=1$.}
	\label{tab4}\centering
	\begin{tabular}{ccccccccccccccc}
		\hline
		{$\beta$} &\,\,\,\,\,&{$0$}&\,\,\,&{$0.1$} &\,\,\,& {$0.3$}&\,\,\,&{$0.5$} &\,\,\,& {$0.7$} &\,\,\,& {$0.9$} &\,\,\,& {$1$}\\
		\hline
		{$r_{\rm GUP}$} &\,\,\,\,\,&{$2.02500$}&\,\,\,&{$2.02188$} &\,\,\,& {$2.01563$}&\,\,\,&{$2.00938$} &\,\,\,& {$2.00313$} &\,\,\,& {$1.99688$} &\,\,\,& {$1.99375$}\\
		{$r_{\rm ph}$} &\,\,\,\,\,&{$3.03750$}&\,\,\,&{$3.03281$} &\,\,\,& {$3.02344$}&\,\,\,&{$3.01406$} &\,\,\,& {$3.00469$} &\,\,\,& {$2.99531$} &\,\,\,& {$2.99063$}\\
		{$b_{\rm c}$} &\,\,\,\,\,&{$5.26110$}&\,\,\,&{$5.25299$} &\,\,\,& {$5.23675$}&\,\,\,&{$5.22051$} &\,\,\,& {$5.20427$} &\,\,\,& {$5.18803$} &\,\,\,& {$5.17991$}\\
		\hline
	\end{tabular}
\end{table}

\begin{figure}[h]
	\centering
	\includegraphics[width=.45\textwidth]{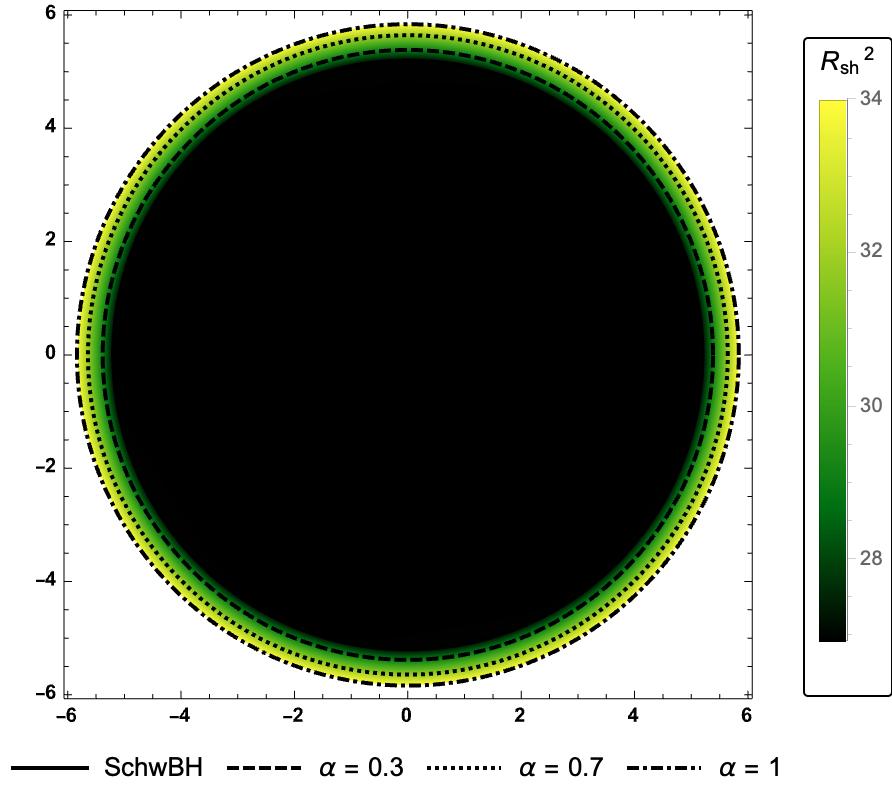}
	\caption{\label{fig8} Visualizing the GCSBH solution shadows from the viewpoint of an observer positioned at spatial infinity, highlighting the changes with various $\alpha$ values while keeping $M=1$ and $\beta = 0.1$.}
\end{figure}

\begin{figure}[h]
	\centering 
	\includegraphics[width=.45\textwidth]{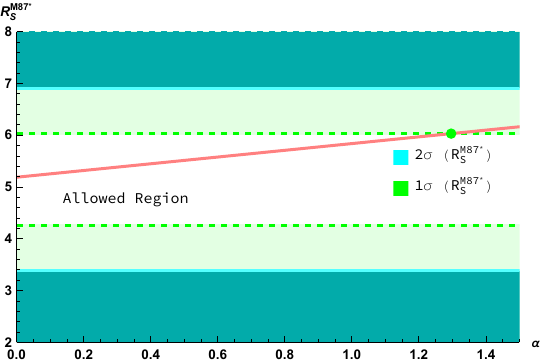}
	\hfill
	\includegraphics[width=.45\textwidth]{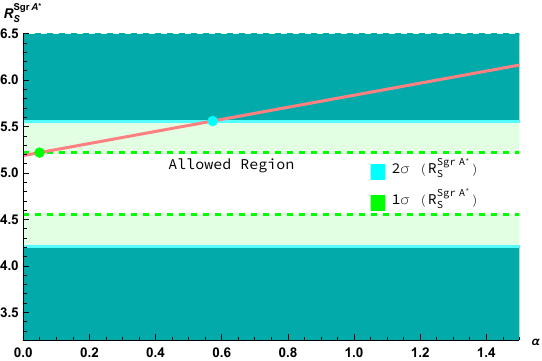}
	\caption{\label{fig9}
		Shadow radius of the MCRBH, depicted in units of the black hole mass $M$ and plotted with respect to the parameter $\alpha$ with $\beta = 0.1$, as characterized by the metric function in Eq. \eqref{MetricFunction1}. The dark cyan regions indicate $\alpha$ values inconsistent with observations of stellar dynamics for M87* (left panel) and Sgr A* (right panel). The white and light green shaded areas correspond to the EHT horizon-scale images of M87* and Sgr A* at $1\sigma$ and $2\sigma$ confidence levels, respectively. In the right panel, these shaded regions represent $\alpha$ values consistent with the averaged Keck and VLTI mass-to-distance ratio priors for Sgr A*.
	}
\end{figure}

It is clear from equation $R_{\text{sh}}= b_{\text{c}} \sqrt{f(r_{o})}$ that the shadow radius depends on the observer position $r_{\rm o}$ \cite{PerlickPR2022,FengEPJC2020,CapozzielloJCAP2023,Capozziello2023,PerlickPRD2018,VagnozziCQG2023,KocherlakotaPRD2021}. We observe that it reduces to $R_{\rm sh}= b_{\rm c}$ for a faraway observer from a black hole with an asymptotically flat metric. This simplification is justified by the fact that, at a considerable distance from the black hole, $f(r_{\rm o}) \approx 1$.
The above condition is satisfied for the Schwarzschild metric when the mass is significantly smaller than the observer's distance $r_{\rm o}$, indicating that the distance between the observer and the black hole is substantially larger than its gravitational radius. Comparing the distances of M87* and Sgr A* from an observer, which are around 16.8 Mpc and 8 kpc, respectively, to their gravitational radii of about $\mathcal{O}(10^{-7})$ pc, we observe that this condition is met in both instances.
We now explore how the GCSBH parameter $\alpha$ affects the size of the shadow radius in spherical symmetry relying on the MCRBH solutions observed by an observer at spatial infinity and on $r_{\rm ph}$. We display the shadows produced by various values of $\alpha$ in Fig. \ref{fig8}, which is associated with the lapse function \eqref{MetricFunction1}.
From Tables \ref{tab3} and \ref{tab4} one can see that the parameter $\alpha$ has a greater impact than the parameter $\beta$ on the behavior of the GCSBH. It considerably influences the variation in shadow size, leading to an expansion of the shadow size as $\alpha$ increases.  In addition, the chosen value of $\alpha = 0.3$ in Fig. \ref{fig8} is within the uncertainty boundaries associated with both M87* and Sgr A* supermassive black holes. The permissible ranges will be calculated based on the data from M87* and Sgr A*.
An intriguing opportunity to conduct an accurate test of gravitational theory in the strong and relativistic field regimes has been presented by the BH shadow reports made by the EHT Collaboration. Applying the Schwarzschild deviation parameter $\delta$ can provide boundaries on the parameters of a particular BH scenario.
Inspired by Refs. \cite{VagnozziCQG2023,KocherlakotaPRD2021} and observational data, we aim to establish constraints on the GCSBH parameter $\alpha$ based on uncertainty and shadow image observations of M87* and Sgr A* provided by the EHT \cite{AkiyamaL12019,AkiyamaL52019,AkiyamaL62019,AkiyamaL122022,AkiyamaL172022,DoScience2019,GravityCollaborationAA2022}.
The observer is located at a distance of 8.277 kpc from Sgr A* and 16.8 kpc from M87*. The masses of the supermassive black holes Sgr A* and M87* are around $4.3 \pm 0.013 \times10^{6}\text{M}_{\odot}$  and $6.5 \pm 0.90 \times10^{9} \text{M}_{\odot}$, respectively.
Given the Schwarzschild shadow deviations of M87* as $\delta_{\text{M87*}} = -0.01\pm 0.17$, where $\frac{R_{\text{S}}}{M} =3\sqrt{3}(1+\delta_{\text{M87*}})$ represents the shadow radius level, the shadow size of M87* is restricted to the range $4.26 \le \frac{R_{\text{S}}}{M} \le 6.03$ within the $1\sigma$ confidence region.
In this way, by averaging Keck and VLTI-based estimates, denoted as $\delta_{\text{Sgr A*}} \simeq 0.060^{+0.065}{-0.065},\text{(Avg)}$, and using $\frac{R{\text{S}}}{M} =3\sqrt{3}(1+\delta_{\text{Sgr A*}})$ to represent the shadow radius, the size of Sgr A*'s shadow is constrained to the range $4.55 \le \frac{R_{\text{S}}}{M} \le 5.22$ within the $1\sigma$ confidence level.
Fig. \ref{fig9} illustrates the variation of the shadow image radius with the parameter $\alpha$ for both M87* and Sgr A*, incorporating uncertainties at the $1\sigma$ level for M87* and at the $1\sigma$ and $2\sigma$ levels for Sgr A*. The numerical values representing the upper bounds in $\alpha$ are provided in Table \ref{tab5}.
This comparison involves the shadow radius of both M87* and Sgr A*, where we meticulously selected appropriate parameter values within the $1\sigma$ and $2\sigma$ uncertainty ranges. Notably, the data for Sgr A* provides a more robust constraint on the parameter $\alpha$ compared to M87*, as there are points that cross the corresponding upper bound of the shadow radius.

\begin{table}[h!]
	\caption{Acceptable values for the GCSBH parameter $\alpha$ can be determined from the curve shown in Fig. \ref{fig9}. These values are associated with the BH shadow radius that aligns with the EHT horizon-scale images of M87* and Sgr A* within the $1\sigma$ and $2\sigma$ confidence levels.}
	\label{tab5}\centering
	\begin{tabular}{cccccc}
		\hline
		{$\alpha$} &\multicolumn{2}{c}{$1\sigma$}& & \multicolumn{2}{c}{$2\sigma$} \\
		\cline{2-3} \cline{5-6}
		
		{} & {Upper} & {Lower} &{}& {Upper} & {Lower} \\
		\hline
		$\text{M87*}$ & -- & $1.30$ & & -- & -- \\
		$\text{Sgr A*}$ & -- & $0.05$ & & -- & $0.57$ \\
		\hline
	\end{tabular}
\end{table}

\section{Conclusion\label{Conc}}
\label{sec:Conclusions}
To sum up, we  considered the GUP  with linear and quadratic moment, and investigated the thermodynamic properties of the Schwarzschild black hole, and obtained the black hole's temperature by the minimal momentum feature of a higher-order GUP. Then, we derived the effective correction of Planck constant by GUP, and calculated the GUP-corrected  Hawking temperature. Furthermore, we investigated the functions of heat capacity and entropy of black holes in the semi-classical framework, and discussed the phase transition caused by GUP effect in detail. In addition, we also analyzed the influence of GUP on Helmholtz free energy
for the Schwarzschild black hole. In this manner, the evaporation process in this context is investigated. For a black hole when the mass goes to zero, the temperature goes to infinity but in the presence of the GUP, we observe that when the mass goes to zero for a temperature goes to zero the mass is remnant mass and the remnant mass is very important because the information of the black hole was not lost.
Also, we see the effect of the GUP parameters on the remnant mass, and when the GUP parameters go to zero the remnant mass goes to zero and the temperature has the Hawking temperature behavior. On the other hand, the importance of the remnant mass can be seen in the evaporation time of the black hole because in the ordinary case, the final mass of the black hole is zero but here the initial mass in the evaporation evolution is $M_i$ but the final mass of the evaporation time is the remnant mass. Also, we observe that for the same initial mass, the evaporation time increases when the remnant mass decreases. Moreover,  we  verified our results within the limits of the HUP which are already found in the literature and we compared all these results with each other especially demonstrating them graphically.

Furthermore, we have derived a static and spherically symmetric GCSBH solution, incorporating quantum corrections from the GUP approach, achieved by replacing the mass with the GUP mass. In this framework, the event horizon radius, photon sphere radius, and critical impact parameter are affected by the GCSBH parameters. Specifically, increasing $\alpha$ and $\beta$ results in an increase and decrease, respectively, in $r_{\rm GUP}$, $r_{\rm ph}$, and $b_{\rm c}$ compared to a Schwarzschild black hole, where $\alpha = 0 =\beta$. We were thus motivated to investigate the radius of the corresponding black hole shadow by examining how the GUP approach influences the critical impact parameter, thereby exploring the astrophysical implications of the GUP formalism. Consequently, we constrained the spacetime parameter $\alpha$ using data from the EHT for both M87* and Sgr A*. Our analysis revealed that Sgr A* imposes a more stringent constraint on the parameter.

\section*{Acknowledgments}
H. Chen is supported by the Doctoral Foundation of Zunyi Normal University of China (BS [2022] 07, QJJ-[2022]-314), S.H. Dong acknowledges partial support of project, S.H. Dong is on leave from IPN due to permission for a research stay in China.
The research of S.Z. was supported by the European Union.-Next Generation UE/MICIU/Plan de Recuperacion, Transformacion y Resiliencia/Junta de Castilla y Leon,
RED2022-134301-T financed by MICIU/AEI/10.13039/501100011033,
and PID2020-113406GB-I00 financed by MICIU/AEI/10.13039/501100011033.
The research was supported by the Long-Term Conceptual Development of a University of Hradec Kr\'alov\'e for 2023, issued by the Ministry of Education, Youth, and Sports of the Czech Republic.
\section*{Data Availability Statement}
We have cited all the data that we have used.




\end{document}